\documentclass{ws-procs10x7}
\usepackage{balance}

\newcommand{\oot}{\overline {126}}

\newcommand{\nnu}{\nonumber\\}
\newcommand{\be}{\begin{equation}}
\newcommand{\ee}{\end{equation}}
\newcommand{\bea}{\begin{eqnarray}}
\newcommand{\eea}{\end{eqnarray}}

\newcolumntype{d}[1]{D{.}{.}{#1}}

\makeindex
\begin{document}

\title{ MSGUT : the next Avtar}

\author{ Charanjit S. Aulakh$^*$  }

\address{Dept. of Physics, Panjab Univ.,Chandigarh, 160014,INDIA
 \\$^*$E-mail:  aulakh@pu.ac.in}




\twocolumn[\maketitle\abstract{ We   report  the viability of the
Type I seesaw mechanism  in 3 generation   renormalizable SO(10)
GUTs   based on  the $\mathbf{{210-10-\oot-126-120}}$ Higgs
system. The $\mathbf{120}$ -plet and $ \mathbf{10}$-plet Higgs fit
charged fermion masses while small $\mathbf{\overline{126}} $-plet
couplings enhance Type I seesaw neutrino masses   to viable values
and make the fit to light fermions accurate.  For the 3 generation
CP conserving case we display accurate charged fermion fits
($\chi^2_{tot} < .2 $)
 which imply Type I neutrino masses   $10^2- 10^3$ times larger than
  the $\mathbf{10-\overline{126} }$ scenario.   The
   correct ratio of neutrino mass squared splitting,
large($\theta_{23}^{PMNS}$) and small($\theta_{13}^{PMNS}$) mixing
angle are achievable.   $\theta_{12}^{PMNS}$ is however small and
  indicates that -as in the Type I $\mathbf{10-\oot}$ case - a
fully realistic  fit to the lepton mixing data also  \emph{
requires} CP violation. }]

\section{Introduction}

The discovery of neutrino oscillations  has driven an intense wave
of research into supersymmetric seesaw mechanisms, Left-right
symmetric models and particularly  the minimal Supersymmetric
Grand Unified Theory\cite{aulmoh} (MSGUT) with Higgs multiplets
$\mathbf{{210-10-\oot-126}}$ . GUT scale Spectra, threshold
effects  and fermion spectrum fits \cite{spectcmb,ffitcmb} have
all received much attention.

 Until last year this `Babu-Mohapatra' (BM) program seemed
  successful and accurate generic  fits to
   all known fermion data  using Type I,
Type II and mixed  seesaw mechanisms were obtained\cite{ffitcmb}.
However it was just assumed that the required overall scale and
relative strength of the Type I and Type II seesaw masses   could
be realized  in appropriate Susy GUTs.   The first
survey\cite{gmblm} of this question in the MSGUT  revealed serious
difficulties in obtaining Type II over Type I dominance and also
in obtaining large enough Type I neutrino masses. Using a
convenient parametrization of the MSGUT spectra and couplings in
terms of the single ``fast'' parameter ($x$) which controls MSGUT
symmetry breaking , as emphasized by the authors of the first
reference in[8], a complete proof was then given\cite{blmdm} of
the failure of the Seesaw mechanism in the context of the MSGUT
and confirmed by another group\cite{domother}. We also
suggested\cite{blmdm,nuMS} a fix for the problem of too small
seesaw masses using an additional $\mathbf{120}$plet. In our
fitting \emph{ansatz} the small $\mathbf{\oot} $ -plet couplings
give appreciable contributions only to light charged fermion
masses {\textit{and}} enhance the Type I seesaw masses  to viable
values since the Type I seesaw masses are \emph{inversely}
proportional to the ${\bf{\oot}}$ coupling. The 2-3 generation
case was first analyzed\cite{nuMS} as a toy model of the dominant
core of the complete hierarchical fermion mass system. Consistency
\emph{required}    $ {\mathbf{m_b-m_s = m_{\tau}-m_{\mu}}} $  at
the GUT scale $M_X$ and predicted near maximal PMNS mixing  for
wide parameter ranges. In the current contribution  we  report on
the extension\cite{msgreb} of our 2 generation toy model to the 3
generation CP conserving case using a procedure based on an
expansion in a Wolfenstein type parameter around the dominant 23
generation core of the fermion hierarchy.

\subsection{  Seesaw Failure   in the MSGUT}

  The Type I and Type II
 seesaw  Majorana masses of  the light neutrinos
 in the MSGUT are  ($ \hat{h},\hat{f}$ are proportional to
$\mathbf{10,\oot}$    Yukawas) :

\bea M_{\nu}^I &=& (1.70 \times 10^{-3} eV) ~ { F_{I}}~
\hat{n}~{\sin \beta}\nnu
 M_{\nu}^{II} &=& (1.70 \times 10^{-3} eV) ~{ F_{II}}
 ~\hat{f}~{\sin \beta}\nnu
    \hat{n}&=& ({\hat h} -3 {\hat f}) {\hat f}^{-1}
    (  {\hat h} -3 {\hat f})\nonumber
     \eea

$F_I,F_{II}$ are   specified functions of $x$.

 Typical BM-Type II fits  require
     $R={F_I/F_{II}}  \leq 10^{-3}$  so it     not be overwhelmed by
   Type I values it implies. Such R values are un-achievable
     in the MSGUT. In Type I fits
      values of $F_I\sim 100$ are needed
     but  are not achievable anywhere over the  complex $x $
     plane\cite{blmdm,domother}.

\subsection{The new $\mathbf{10-120-\oot}$ scenario }

    To resolve the difficulty
 we proposed\cite{blmdm,nuMS}
    that   $\mathbf{\oot}$ couplings be reduced
     much below the level where they are important for 2-3 generation
    masses  and    introduced $\mathbf{120}$ plet  for
    charged fermion mass fitting (previously accomplished by ${\bf{\oot}}$
  couplings  comparable to those of the ${\bf{10}}$-plet).
Fermion masses in such GUTs are( $m$ :~Dirac, $M$: Majorana):
\bea \hat m^u &=&  v( {\hat h} + {\hat f} + {\hat g} )\nnu
 \hat m_{\nu}&=&  v ({\hat h} -3 {\hat f}  + r_5' {\hat{g}})
 \nnu
\hat m^d &=& { v (r_1} {\hat h} + { r_2} {\hat f}  +
r_6 {\hat g}) \\
   \hat m^l &=&{ v( r_1} {\hat h} - 3 {  r_2} {\hat f} + r_7{\hat g})
       \label{120mdir}\nnu
M_{\nu}^{I} &=& v r_4 \hat{n} \nnu \hat{n}&=& ({\hat h} -3 \hat{f}
- r_5' {\hat g}) {\hat f}^{-1}
    (  {\hat h} -3 \hat{f }+ r_5'{\hat g})\nonumber\eea

\noindent  where   $r_i$ are coefficients fixed by the Susy GUT.
  We assume CP conservation i.e  fermion  Yukawa couplings and coefficients $r_i$
are both real.  Note that with only $r_i$ complex and the Yukawa
couplings real (i.e spontaneous CP violation) the resulting NMSGUT
has only 12 Yukawa coupling parameters  i.e 3 less than in the
MSGUT with complex Yukawas !

Matching    mass terms above    to
 renormalized MSSM mass matrices  at $M_X$ introduces
   unitary matrices (specifying the MSSM$\subset$MSGUT embedding)
   which  are of vital relevance  for
    calculation of   exotic signatures\cite{gmblm} :
\bea {\hat m}^u &=&
  V_u^T  {D}_u Q\nnu
 {\hat m}^d &=&  V_d^T  {D}_d R \\
{\hat m}^l &=&  V_l^T  {D}_l L\nonumber
   \eea
  $D_{u,d,l}$ are   masses at $M_X$  and
$V_u,Q,V_d,R=C^T Q,L,V_l$  unitary matrices ($C$ is the CKM
matrix).
  Rewriting $V_{u,d,l}$  : $ V_d = \Phi_d R \hspace{2mm};
   \hspace{2mm} V_u = \Phi_u  Q \hspace{2mm};
   \hspace{2mm} V_l = \Phi_l L $ and
separating symmetric and antisymmetric parts  :$~ Z = \Phi^T {D} +
{D} \Phi \hspace{2mm}; \hspace{2mm} A = \Phi^T D - D \Phi $
  allows the problem to be reduced to that of
determining the     matrices $\Phi_{u,d,l}, {\cal D} = R L^T $ and
coefficient $\hat{r}_2$ such that : \bea
   \hat S_3 & = & \hat S_X
     + {\hat r_2}  {\frac{Tr Z_d}{Tr Z_u}}
 ( {\hat S_2}  - {\frac{(Tr Z_d - Tr Z_l)}{ Tr Z_d}}{\hat X}) \nnu
\hat{A_1} &=&  C  \hat A_d C^T - {\hat A}_u =0 \nnu
 {\hat A}_2^{\pm} &=& \hat A_d  ~ \mp ~ {\cal  D } {\hat A}_l   {\cal D
 }^T=0\nnu
 \hat X &\equiv& {\frac{3 Z_d +{\cal D} Z_l {\cal D}^T }{3 Tr Z_d +
Tr Z_l}}\\ \hat S_X &\equiv& \hat X - C^T {\frac {Z_u}{Tr Z_u}} C
\nnu {\hat A}_{u,d,l} &\equiv& { {A_{u,d,l}}/ {\sqrt{{\vec
A_{u,d,l}} ^2} }}
 \nonumber \label{sumrule} \eea

 \noindent the ambiguous sign  above corresponds to
the   branch of the 2 generation model that one is expanding about
i.e   $\hat{A_2}^{\pm}=0$   for $\chi_l=\pm \bar\chi_l$.The
definitions are such that   $\hat S_2=0\Rightarrow $   $\hat S_3=0
$ reduces to $\hat S_1=0$.

  These  equations are  too complicated
to solve analytically in the 3 generation case. Our
approach\cite{nuMS,msgreb} is to consider the 23 sector as the
dominant ``core'' of the fermion mass hierarchy and   to expand
around it. With our reality assumption all unitary matrices can be
given the parametrization($O_{ij}$ are orthogonal rotations) :
 $ O   =   O_{23}(\chi) \cdot O_{13}(\phi) \cdot O_{12}(\theta) $.
 In view of the encouraging
results of our 2 generation model for the 23 dominated fermion
hierarchy,  we looked for solutions  as an expansion (in powers of
$\epsilon\sim \sqrt{\theta_{23}}\sim \theta_{12}\sim .2 $) around
the 2 generation results.  The 2 generation case gave\cite{nuMS}
$\hat S_2\sim O(\epsilon^3)$   and  $~~ \hat f = {\frac{1}{2v}}
{\hat r_2} Tr Z_d {\hat S_2}$.
 The $O(10^{-2})$ suppression provided by the ratio
$d_3/v$     implies   that $(\hat f)\sim .01\epsilon^{3+\delta}$
when    ${\hat r_2}\sim \epsilon^{\delta}$. This ensures the
enhancement of Type I neutrino masses that is the rationale for
this fitting scenario.

 We expand  all the
 orthogonal matrix angles  $(\theta , \phi ,\chi)_{u,d,l,{\cal D}}$
  as well as the free parameter $\hat r_2$ in powers of  $\epsilon
 $ and take $ \chi_d^{(0)}=\chi_u^{(0)}=\chi_l^{(0)}=\chi_{\cal D}^{(0)}
 $   corresponding to the $\chi_l=+\bar\chi_l $
 solution,   \emph{or }
 $ \chi_d^{(0)}=\chi_u^{(0)}=-\chi_l^{(0)}=-\chi_{\cal
  D}^{(0)}$    corresponding to the $\chi_l=-\bar\chi_l $
  solution\cite{nuMS}.

  Note that the the CKM matrix drives the off-diagonality
   of  $\hat   S_3=0$    and since  $\phi_c\sim \epsilon^3$
  the angles $\phi_{u,d,l,{\cal D}}$ are $O(\epsilon)$ or
     smaller, so we incorporate this  behaviour of the angles
     $\phi_{u,d,l,{\cal{D}}}$ into our \emph{ansatz}
     from the beginning : resulting in a considerable simplification of the
     ansatz.
   Details   may be found in  \cite{nuMS,msgreb} but two points are
crucial : the value of $\chi_d^0$ and the constraint $d_3= l_3-l_2
+ d_2 $ are still fixed by the expansion upto $O(\epsilon^2)$ just
as in the toy model.
 A simple criterion is to   define    $\chi^2 $ functions
measuring the deviation    from central  data and  terminating the
perturbation expansion   when  these are   small. For the central
quark masses and angles we used Das-Parida\cite{dasparida} central
values (for $\tan\beta(M_S) =55$ and at $M_X=2\times 10^{16} GeV$
) except for $d_3$ as explained above .

 The Type I seesaw mass formula   is
\bea \hat M_{\nu}^I&\simeq & \frac{v^2}{2\widehat{\bar{\sigma}} }
(\widehat{h}+ r_5^{'}\widehat{g}-3 \hat f)^T
\hat{f}^{-1}(\widehat{h}+ r_5{'}
\widehat{g} -3 \hat f)\nonumber\\
  &\equiv&(1.70 \times 10^{-3} eV) R^T \hat n R F_I \sin \beta
\nonumber\\
&=&  {L}^T \mathcal{P} D_{\nu}\mathcal{P}^T  {L} \label{TypeI}\\
{\cal{P}}& =& D^{\dagger} {\cal{N}} \qquad ;\qquad
{\frac{{m}_{sol}^2}{{m}_{atm}^2}}={\big|}{\frac{\hat{n}_{\mu}^2
-\hat{n}_e^2}{\hat{n}_{\mu}^2 -\hat{n}_\tau^2}}{\big|}\nonumber
\eea
 ${\cal{P}}$ is the  PMNS
matrix   and  $D_{\nu}$ the light neutrino masses
 at $M_X$. ${\cal{N}}$ diagonalizes $\hat n : \hat n= {\cal{N}}\hat
n_{diag} {\cal{N}}^T$.

\section{Examples of  3 generation  fits}

We have obtained a number of examples of acceptable fits -modulo
the assumption that CP violation is neglected- by going to high
orders in $\epsilon$ (sometimes as high as $O(\epsilon^{20})$).
For reasons of space we confine ourselves to quoting the values
found for one such fit: $ {\hat r}_1  = 15.27~  \mathbf{;}~  {\hat
r}_2 =0.255 ~ \mathbf{;} ~ r_6=0.0187 ~ \mathbf{;}  ~
r_7=0.023305$. \bea {\hat{h}}  &=& \left(\begin{array}{ccc}
 -0.000299&0.000039 &-0.02227 \\ 0.000039 &
    0.00929 &-0.13482\\-0.02227&-0.13482&0.481333
  \end{array} \right)\nnu
 \nnu{\hat{f}}  &=&
\left(\begin{array}{ccc}0.0000132&0.0000277&-0.0001\\0.0000277&-0.0000146&
    0.000057\\-0.0001&0.000057&0.000040
  \end{array} \right) ~~ \\
 {\hat{g}}  &=& \left(\begin{array}{ccc}0& 0.001644& -0.02478 \\
 -0.00164 &
      0.& -0.119710\\0.02478&0.119710&0
  \end{array} \right)\nonumber\eea
The eigenvalues of $\hat h,\hat f,\hat g$ are \bea \hat h &:& ~
~0.52~~\quad;~~0.028~~\quad;~~1.92\times 10^{-5} \nnu \hat f &:&
1.33\times 10^{-4} ~~;~~1.11\times 10^{-4}~~;~~0.17\times
10^{-4}\nnu \hat g&:& \pm 0.122 ~~;~~0\eea
  The premises of our\cite{blmdm,nuMS,msgreb} scenario are indeed respected.
Then the reconstructed values of the charged fermion masses are(in
GeV) : $M_U  =  \{95.148,0.211,0.00077\} \mathbf{;}$ $
 M_D  = \{1.584,0.0298,0.0015\} \mathbf{;}$ $
M_l =  \{1.629,0.0753,0.00036\}$ and  the  CKM angle magnitudes
:$~\theta_{12} =0.227  ~\mathbf{;}~
    \theta_{13} =0.00216  \mathbf{ ;} ~   \theta_{23} =0.038  $.
The charged fermion fit $\chi^2$ values are $ \chi^2_m= 0.126
 ~ \mathbf{;}  ~ \chi^2_{CKM}=.016  ~\mathbf{;}
 ~\chi^2_{tot}= 0.142 $.

 Fixing  $r_5'$ by imposing  $ R_{ijik}=
  \big|{\frac{{\hat n}_i^2 -{\hat n}_j^2}{{\hat
n}_i^2-{\hat n}_k^2}}| =.32 $  gives 6 solutions which are however
3 closely related pairs. One finds that in only   one  case is
  $\theta_{23} $  large and  $ \theta_{13} $   reasonably small.
   \emph{However the value of the 12 sector mixing is very
low}.  On the other hand the large ( about 200 times larger than
the Type I fits in the $ \mathbf{{10-\oot}}$ scenario : see below)
value of the largest $\hat n$ eigenvalue together with the
satisfactory value of the mass squared splitting ratio means that
the problem with too small neutrino masses is unlikely to appear
even for generic values of the GUT  scale breaking, leave alone
regions where the coefficient function $F_I$ is itself large.

 In the  almost viable case
$r_5'=-.333  ~;$ \hfil\break $ \phi_{\cal D}=0.053  ~;~
\theta_{\cal D}= 0.435 ~;~ \chi_{\cal D}= 0.0257 $ and $\hat n_1 =
113.8
 ~;~ n_2 =20.045   ~;~ n_3 = 1.97\times 10^{-5},  $ while $
\sin^2 \theta_{12}^{PMNS} =.073  ~;~ \sin^2\theta_{23}^{PMNS} =
.77  ~;~\theta_{13}^{PMNS}=.176 $.

\section{Discussion, Conclusions and Outlook}
  We have reported progress  towards
   a completely realistic fit of all known
charged fermion and neutrino mass data using the mass relations
and RG evolution common to any SO(10) Susy GUT with a
$\mathbf{{10-120-\oot}}$  FM Higgs system. Specifically, we have
shown that in the quasi realistic 3 generation but CP
preserving(real) case we are able to obtain accurate charged
fermion fits, neutrino mass parameters and a PMNS mixing pattern
that can be large  in the 23 sector and small in the 13 sector.
The remaining deficiencies, namely   simultaneous large mixing in
the 12 sector and the fit of the MSSM CKM CP phase in the first
quadrant can presumably be remedied in the complex 3 generation
case, in close analogy with the  ${\bf{10-\oot}}$  case where a
successful Type I fit could be found\cite{ffitcmb} \emph{only}
when  CP violation was introduced. Very recently another group has
implemented our scenario in the $\mathbf{120}$ extended MSGUT with
spontaneous CP violation and an  ad-hoc $Z_2$ symmetry imposed to
improve tractability. Using the ``downhill simplex'' method of non
linear fitting they obtain a very accurate and realistic 3
generation fit\cite{grimusnu}. With the completion of this program
we  will be in possession of a well defined Susy GUT compatible
with all low energy data as well as information on the embedding
of the MSSM in the MSGUT  ( coded in the Unitary matrices
$\Phi_{u,d,l},{\cal D}$ ) which emerges as the most valuable
corollary product of the fitting procedure\cite{gmblm}. It is only
then that we will be able to enter the third phase of the GUT
program in which the exotic process ($\Delta B \neq 0$, LFV etc )
predictions will finally be linked sufficiently tightly to low
energy data  as to make the
 search for exotic processes  a falsifiability test rather than a hopeful
check on a lottery bet.

\end{document}